\documentclass[letterpaper,10pt,conference]{ieeeconf} 

\IEEEoverridecommandlockouts
\usepackage{graphicx}

\usepackage{amsmath}
\usepackage{amssymb}
\usepackage{amsfonts}
\usepackage{mathtools}
\usepackage{MnSymbol}
\usepackage{subcaption}
\usepackage{amsthm}
\usepackage{algpseudocode}
\usepackage[usenames,dvipsnames,svgnames,table]{xcolor}
\usepackage{float}
\usepackage{caption}
\floatstyle{plaintop}
\restylefloat{table}
\usepackage[vlined,ruled,linesnumbered]{algorithm2e}
\newtheorem{definition}{Definition}

\newtheorem{problem}{Problem}

\newtheorem{example}{Example}
\usepackage{multirow}
\newcommand{\levent}{\lozenge}

\newcommand{\luntil}{\mathcal{U}}
\newcommand{\lnext}{\bigcirc}
\newcommand{\TS}{\mathcal{T}}
\newcommand{\FA}{\mathcal{A}}

\renewcommand{\baselinestretch}{0.97}
\usepackage{eso-pic}
\newcommand\AtPageUpperMyright[1]{\AtPageUpperLeft{
 \put(\LenToUnit{0.5\paperwidth},\LenToUnit{-1cm}){
     \parbox{0.5\textwidth}{\raggedleft\fontsize{9}{11}\selectfont #1}}
 }}
\newcommand{\conf}[1]{
\AddToShipoutPictureBG*{
\AtPageUpperMyright{#1}
}
}
\title{\LARGE \bf
Fair Planning for Mobility-on-Demand with Temporal Logic Requests
}

\author{Kaier Liang and Cristian-Ioan Vasile
\thanks{Kaier Liang and Cristian-Ioan Vasile  are with the Mechanical Engineering and Mechanics Department at Lehigh University, PA, USA: {\tt\small \{kal221, cvr519\}@lehigh.edu}}        
}     
\conf{2022 IEEE/RSJ International Conference on Intelligent Robots and Systems (IROS) 23–27, October, 2022, Kyoto, Japan} 

\begin{document}


\maketitle
\thispagestyle{empty}
\pagestyle{empty}

\begin{abstract}

Mobility-on-demand systems are transforming the way we think about the transportation of people and goods.
Most research effort has been placed on scalability issues for systems with a large number of agents and simple pick-up/drop-off demands.
In this paper, we consider fair multi-vehicle route planning with streams of complex, temporal logic transportation demands.
We consider an approximately envy-free fair allocation of demands to limited-capacity vehicles based on agents' accumulated utility over a finite time horizon, representing for example monetary reward or utilization level.
We propose a scalable approach based on the construction of assignment graphs that relate agents to routes and demands, and pose the problem as an Integer Linear Program (ILP).
Routes for assignments are computed using automata-based methods for each vehicle and demands sets of size at most the capacity of the vehicle while taking into account their pick-up wait time and delay tolerances.
In addition, we integrate utility-based weights in the assignment graph and ILP to ensure approximative fair allocation.
We demonstrate the computational and operational performance of our methods in ride-sharing case studies over a large environment in mid-Manhattan and Linear Temporal Logic demands with stochastic arrival times.
We show that our method significantly decreases the utility deviation between agents and the vacancy rate.
\end{abstract}

\section{INTRODUCTION}

With the development of urbanization, the demand for transporting people and goods is expanding. Yet simply increasing the number of private vehicles is inefficient for road traffic and not environmental-friendly. On the other hand, the mobility-on-demand system can be economical and sustainable. This system allows passengers to specify their demands and employ a large scale of ride-sharing on the road map, thus reducing the traveling cost, alleviating the traffic congestion and emission \cite{teubner2015economics,caulfield2009estimating,agatz2012optimization}.
However, less attention has been paid to how fair transportation requests are distributed to drivers in mobility-on-demand systems.

A wealth of research has investigated the system and has focused on the real-time route planning and scalability issues with a large number of agents.
Rebalance policies for the congestion and high demand were studied in \cite{pavone2012robotic,wen2017rebalancing,wallar2018vehicle}.
In mesoscopic optimization, using the estimations of traffic congestion, joint operations for autonomous vehicles fleet was studied in \cite{salazar2019congestion,salazar2019intermodal}.
From a microscopic perspective, the requests assignment with a defined cost can be formulated as an optimization problem \cite{santos2015taxi}. One approach is to construct a shareability graph between vehicles and requests for the ride-sharing\cite{santi2014quantifying,7264299}. Based on this approach,  Alonso-Mora et al. create a Requests-Trip-Vehicle (RTV) assignment graph~\cite{alonso2017demand}.
The large ride-sharing problem is encoded using integer linear programming (ILP) and solved almost in real-time. 

However, among the studies above, the requests mainly consisted of simple atomic tasks, such as driving from point A (pick-up location) to point B (drop-off location).
This leaves an unexploited scenario where the requests are complex. For example, a customer may want to purchase a gift from store A or store B, while another customer wants dinner at a restaurant near store A. Suppose both of them have close pick-up positions and send out the requests at a similar time; one vehicle may be able to accommodate both of them by driving to the restaurant and store A, should the waiting and delay times be acceptable for them.
This kind of request can be represented using linear temporal logic (LTL), which is employed in \cite{tumova2016least} for a single-vehicle routing to tackle complex requests assignment.
The map and vehicles are modeled as weighted transition systems (WTS), and the demands are formulated using co-safe LTL (scLTL). Then we graph search algorithms check the ride-sharing feasibility and make the assignment based on the defined cost function. 

Furthermore, research into ride-sharing has often focused on the customer side. The objective of requests assignment and route planning is to minimize travel costs and fairness is usually considered from the customers' perspective\cite{foti2017nash,cao2021optimization}. However, drivers' preferences may not agree with the assignment they received.
Moreover, the demand for drivers may exceed the number available, e.g., during peak hours, which can give drivers an edge in the request-driver relationship, as drivers can have more choice.
Also, there might be some vacant vehicles in the assignment when there is less demand for vehicles during off-peak. Therefore,  fairness should also be considered from the drivers' perspective when allocating requests to tackle the utility disparity among drivers. There are different criteria to judge the fairness, for example, maximizing the minimum utility for vehicles \cite{wolfson2017fairness,lesmana2019balancing}. 

\begin{figure}
  \centering
  \includegraphics[width=0.9\linewidth]{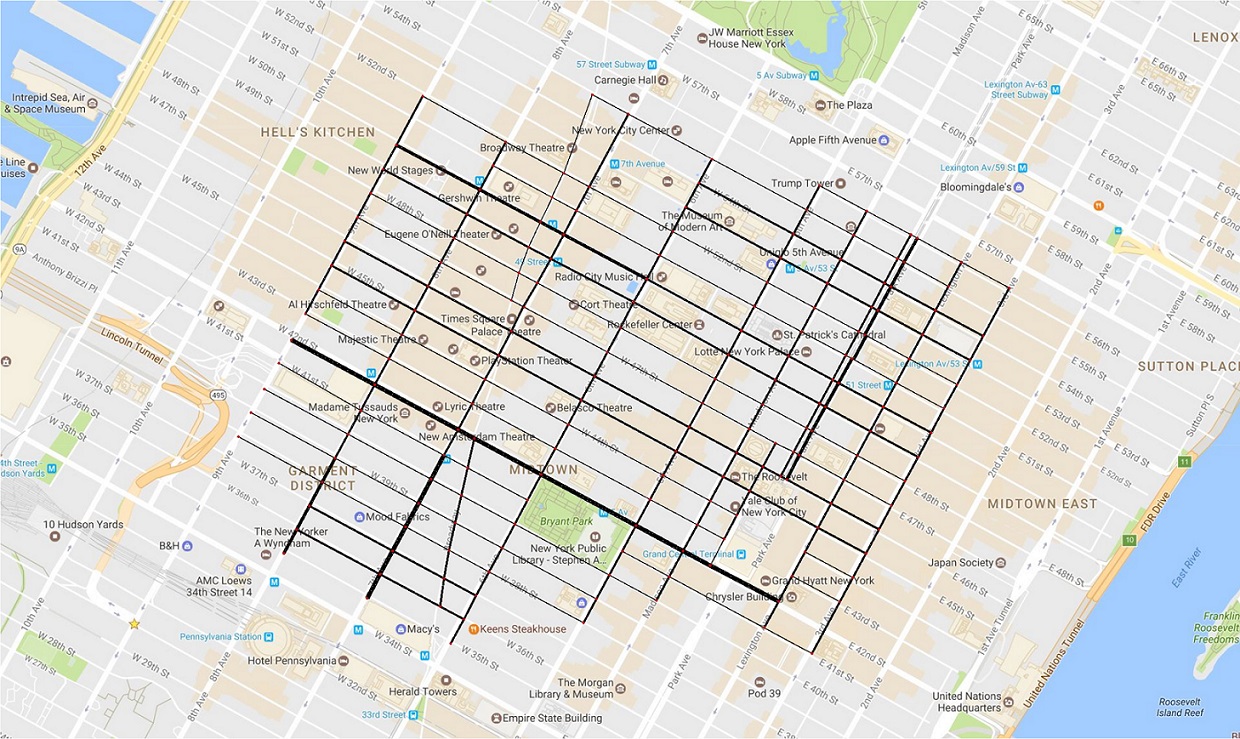}
  \caption{The road network corresponding to part of mid-Manhattan is shown.
  Travel duration estimates are inferred from real taxi travel data in hourly increments \cite{alonso2017demand}.}
  \label{fig:manhattan-network}
\end{figure}

The contributions of this work are the following 1) we propose a multi-vehicle routing problem with fairness constraints on the assignment of temporal logic demands; the arrival time of demands is a priori unknown; we consider fairness in sequential decision making. 2) we propose a combined automata and ILP-based approach that decomposes the problem into a set of small routing problems with scLTL specifications. 3) we propose a weighting scheme for the assignment graph that corrects the history of utility collected by vehicles. 4)  we show the performance of our approach in case studies on part of mid-Manhattan (Fig.~\ref{fig:manhattan-network}); and we show that our approach significantly reduces the deviation of collected utility between vehicles, and the vacancy rate with respect to baseline without fairness considerations.



\section{Preliminaries}
In this section, we introduce the notation used in the paper and review concepts in formal language and automata theory.

We denote the set of real and integer numbers as $\mathbb{R}$ and $\mathbb{Z}$.
The real and integer numbers greater than $a$ are denoted by $\mathbb{R}_{> a}$ and $\mathbb{Z}_{> a}$.
Similarly, we have $\mathbb{R}_{\geq a}$ and $\mathbb{Z}_{\geq a}$ for real and integer numbers greater or equal than $a$. 
For a finite set $S$, we denote its cardinality and the power set as $|\mathrm{S}|$ and $2^S$. 

\begin{definition}[Finite Automaton]
\label{def:dfa}
A deterministic finite state automaton (DFA) is a tuple $\FA=\left(Q_\FA, q_{init}^\FA, 2^\Pi, \delta_\FA, F_\FA\right)$, where $Q_\FA$ is a finite set of states; $q_{init}^\FA \in Q$ is the initial state; $2^\Pi$ is the input alphabet; $\delta_\FA : Q_\FA \times 2^\Pi \to Q_\FA$ is a transition function; $F_\FA \subseteq Q_\FA$ is the set of accepting states. 
\end{definition}

An input word $\boldsymbol{\sigma}=\sigma_{0} \sigma_{1} \ldots \sigma_{n}$ over alphabet $2^\Pi$
generates the \emph{trajectory} of the DFA
$\mathbf{q}=q_{0} q_{1} \ldots q_{n}$
with $q_{init} = q_0$ and $q_{k+1} = \delta_\FA(q_k, \sigma_k)$, for all $k \in \{0,\ldots, n-1\}$.
The trajectory $\mathbf{q}$ is called \emph{accepting} if $q_n \in F_\FA$.

\begin{definition}[scLTL]
\label{def:scltl}
A co-safe Linear Temporal Logic (scLTL) formula over a set of atomic propositions $\Pi$ is defined recursively as:
\begin{equation*}
\phi::=\pi \mid \lnot \pi \mid \phi_1 \lor \phi_2 \mid \phi_1 \land \phi_2 \mid \lnext \phi\mid \phi_1 \luntil \phi_2 \mid \levent \phi,
\end{equation*}
where $\phi_1, \phi_2$ are scLTL formula,
$\pi \in \Pi$ is an atomic proposition,
$\lnot$ (negation), $\land$ (disjunction), and $\lor$ (conjunction) are Boolean operators, and $\luntil$ (until), $\lnext$ (next), and $\levent$ (eventually) are temporal operators.
\end{definition}

The semantics of scLTL formulae are defined over infinite words with symbols from $2^{\Pi}$.
Intuitively, $\lnext \phi$ holds if $\phi$ is true at the next position in the word; $\phi_1 \luntil \phi_2$ expresses that $\phi_1$ is true until $\phi_2$ becomes true; and $\levent \phi$ expresses that $\phi$ becomes true at some future position in the word.
The formal definition of the semantics can be found in~\cite{baier2008modelchecking}.
Given a word $\boldsymbol{\sigma}$ over the alphabet $2^\Pi$ that satisfies the scLTL formula $\phi$,
we denote the satisfaction as $\boldsymbol{\sigma} \models \phi$.
A finite word $\boldsymbol{\sigma}$ satisfies scLTL formula $\phi$ if for all infinite $\boldsymbol{\sigma}'$ the concatenated (infinite) word $\boldsymbol{\sigma}\boldsymbol{\sigma}' \models \phi$.
The finite word $\boldsymbol{\sigma}$ is \emph{minimal} if none of its prefixes satisfies $\phi$.

scLTL formulae can be translated to DFAs using off-the-shelf tools such as scheck~\cite{Latvala03} and spot~\cite{Duret.13.atva}.

\begin{definition}[Weighted Transition System]
\label{def:wst}
A weighted transition system (WTS) is a tuple $\TS=\left(S, s_{\text {init }}, D, W, \Pi, L\right)$, where $S$ is a finite set of states, $s_{init} \in S$ is the initial state, $D \subseteq S \times S$ is a transition function, $W: D \to \mathbb{R}_{+}$ is a weight function, $\Pi$ is a set of atomic propositions and $L: D \to 2^{\Pi}$ is a labeling function.
\end{definition}

The transition from the current state $s$ at time $t$ to the next state $s'$ is reached at time $t' = t + W((s, s'))$ if $(s, s') \in D$.
A \emph{trajectory} of $\TS$ is a finite sequence
$\mathbf{s} = s_0 s_1 \ldots s_n$, such that $s_0 = s_{init}$, and $(s_k, s_{k+1}) \in D$ for all $k \in \{0,\ldots, n-1\}$.
The length of the trajectory $\mathbf{s}$ is $n$,
and its total duration is $W(\mathbf{s}) = \sum_{i=0}^{n - 1} W((s_i, s_{i+1}))$.
The \emph{output trajectory} induced by $\mathbf{s}$
is $\mathbf{o} = L(s_0) L(s_1) \ldots L(s_n)$.
A finite trajectory $\mathbf{s}$ satisfies a scLTL formula $\phi$, denoted $\mathbf{s} \models \phi$, if the induced output trajectory $\mathbf{o}$ satisfies $\phi$.

\section{Problem Formulation}
\label{sec:problem-formulation}

In this section, we formulate the fair mobility-on-demand problem with requests expressed as scLTL specifications and vehicle sharing.
Our goal is to compute assignments of sequentially incoming scLTL requests to a fleet of vehicles such that the total traveling cost is minimized, and fairness among the drivers over the planning horizon is ensured.

\subsection{Vehicle, Environment, and Request Models}
\label{sec:models}

Consider a fleet of vehicles $\mathcal{V} = \{v_1, v_2, \ldots, v_p\}$ deployed
in a road network with intersections $S$
and roads $D \subseteq S \times S$,
where $(s, s') \in D$ represents a road from intersection $s$ to $s'$.
The initial position of vehicle $v \in \mathcal{V}$ is $s_{init, v} \in S$.
All vehicles evolve in discrete time $t\in \mathbb{Z}_{\geq 0}$ synchronized via a global clock.
The traversal duration of road $(s, s')$ is $W((s, s')) \in \mathbb{Z}_{>0}$.

Vehicles are tasked with satisfying a finite set of request $\mathcal{R} = \{r_1, r_2, \ldots, r_m\}$
that arrive sequentially over the horizon time $H \in \mathbb{Z}_{>0}$.
A request $r \in \mathcal{R}$ is defined as a tuple
$r = (\pi_{pick, r}, \phi_r, t_{req, r}, \rho_r, \Omega_{\max, r}, \Delta_{\max, r})$, where
\begin{itemize}
    \item $\pi_{pick, r}$ is a proposition marking the pick-up location;
    \item $\phi_r$ is the scLTL formula specifying the request;
    \item $t_{req, r} \in \{0, \ldots, H\}$ is the request's arrival time;
    \item $\rho_r \in \mathbb{Z}_{>0}$ is the number of required seats;
    \item $\Omega_{\max, r} \in \mathbb{Z}_{>0}$ is the maximum waiting time, i.e., the latest accepted pick-up time is $t_{req, r} + \Omega_{\max, r}$;
    \item $\Delta_{max, r} \in \mathbb{Z}_{>0}$ is the maximum allowed delay.
\end{itemize}

Vehicles have limited transportation capacities.
We denote by $Cap_v \in \mathbb{Z}_{>0}$ and $c_v(t) \in \{0, \ldots, Cap_v\}$
the maximum capacity
and the \emph{available capacity} at time $t$
for vehicle $v \in \mathcal{V}$.
A vehicle $v$ is said to be \emph{available} at time $t$ if $c_v(t) > 0$,
otherwise it is \emph{occupied}, i.e., $c_v(t) = 0$.
The set of available vehicles at time $t$ is denoted by $\mathcal{V}^a_t$.

A group of vehicles $V \subseteq \mathcal{V}$ completes a request $r \in \mathcal{R}$
if they pick up $r$
at the intersection marked with $\pi_{pick, r}$
such that their overall available capacity
is greater than $\rho_r$.
Formally, we have $\mathbf{s}_v \models \tilde{\phi}_r$,
vehicle $v$ is available at time $t_{pick, r, v}$ for all vehicles $v \in V$,
and $\sum_{v \in V} c_v(t_{pick, r, v}) \geq \rho_r$,
where $\tilde{\phi}_r = \levent (\pi_{pick, r} \land \phi_r)$,
$\mathbf{s}_v$ is the finite trajectory of $v$ and
$t_{pick, r, v}$ is the pick-up time for 
$r$ by $v$.
Note that we do not require all vehicles $V$ to pick up their share of request $r$ at the same time.

The delay $\Delta_r$ is the difference between the actual and optimal satisfaction duration.
Formally, $\Delta_r = \max_{v\in V} t_{drop,r,v} - t_{req, r} - t^*_r$, 
where $t_{drop,r,v}$ is the drop off time of request $r$ by vehicle $v$.
($\mathbf{s}_v(0{:}t_{drop,r,v})$ is a minimal satisfying word for $\tilde{\phi}_r$) and
$t^*_r$ is the optimal satisfaction time, i.e., the amount of travel time if a vehicle picks up the request at $t = t_{req,r}$ and not share with other requests.
We require that $\Delta_r \leq \Delta_{\max, r}$.

At current time $t \in \mathbb{Z}_{\geq 0}$, a request is \emph{active} if $t_{req} \leq t$ and it has not been picked-up yet;
a request is \emph{in progress} if it has been picked-up and not completed.
The sets of active and in progress requests at time $t$
are $\mathcal{R}^a_t$ and $\mathcal{R}^p_t$, respectively.

\begin{figure}[htb]
    \centering
    \includegraphics[width = 0.9\linewidth,trim={0cm 0.75cm 0cm 0.85cm},clip]{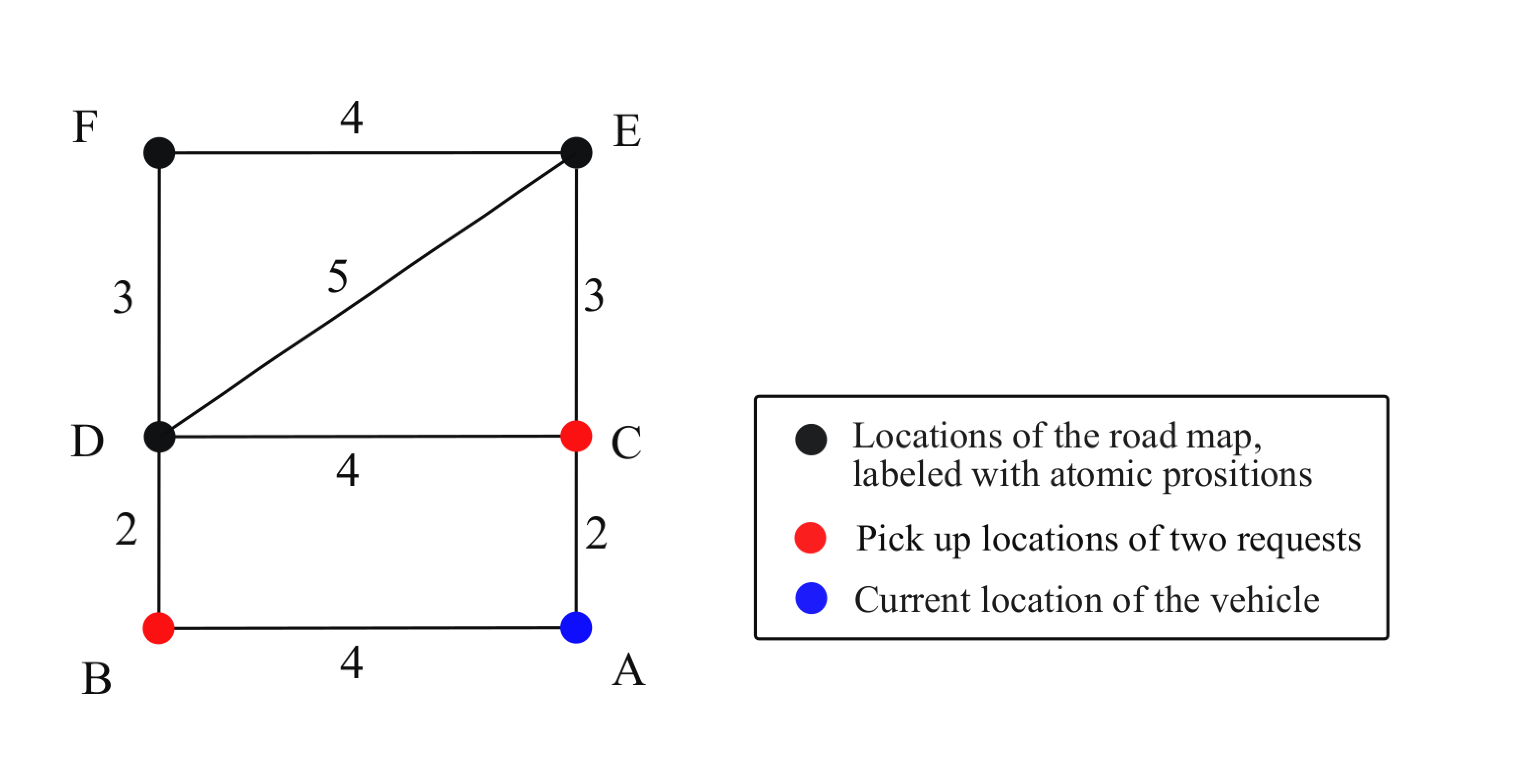}
    \caption{Example of road map model as a WTS, the drivable paths are labeled with weights as traveling cost between each connected nodes.
    For an empty vehicle at location $A$, assuming both requests are active and the maximum delay and waiting time are satisfied, the route of the vehicle $v_1$ with minimal travel cost is $A \rightarrow C \rightarrow D \rightarrow B \rightarrow D \rightarrow F \rightarrow E$.}
    \label{fig:small_road}
\end{figure}

\begin{example}
A small road map in WTS form with the set of atomic proposition $\Pi = \{A, \ldots, F\}$ is depicted in Fig.~\ref{fig:small_road}.
The pick-up locations $\pi_{pick, 1}$ and $\pi_{pick, 2}$ shown in red dots at $C$ and $B$ represent requests with scLTL formulas specifying as $\phi_1=\levent(D \wedge \levent E)$ and $\phi_2=\levent (D \wedge \levent F)$, respectively. 
The blue dot represents the initial position of an empty vehicle $v_1$. 
The least travel times
are $t^*_1 = 9$ and $t^*_2 = 5$ and the assignment planning result is shown in Table.~\ref{tab:example}. 
\end{example}

\begin{table*}
\caption{Example for requests in Figure.~\ref{fig:small_road}}
    \label{tab:example}
\centering
\begin{tabular}{ccccccc} 
\hline
& Pick-up Location & scLTL spec & Arrival Time & Pick-up time & Drop-off Time & Delay\\
\hline
$r_1$ & $s_{\text {pick }, 1} = C$ & $\levent (D \wedge \levent E)$ & $t_{req,1}=0$ & $t_{pick, 1}=2$ & 17 & 8 \\
$r_{2}$ & $s_{\text {pick }, 2} = B$ & $\levent (D \wedge \levent F)$ & $t_{req,2}=0$ & $t_{pick, 2}=8$ & 13 & 8\\
\hline
\end{tabular}
\end{table*}

An assignment $Asg_t: \mathcal{R}^a_t \to 2^{\mathcal{V}^a}$ at time $t = t_{req, r}$
allocates active requests to vehicles when request $r$ arrives.
If the assignment $Asg_t(r) = \emptyset$, then $r$ is \emph{unassigned} at time $t$.
In case this holds for all $t\in \{t_{req, t}, \ldots, t_{req, r} + \Omega_{max,r}\}$, $r$ is unassigned.
An assignment for $r$ may involve multiple vehicles, i.e., $|Asg_t(r)| > 1$.
Requests that are \emph{in progress} cannot be reassigned and vehicles need to be \emph{available} before picking up new requests.
Between request arrivals times, i.e., $t\neq t_{req, r}$, assignments do not change.

The total cost for all requests is defined as
\begin{equation}
\label{eq:total-cost}
    J(\{Asg_t\}_{t=0}^H, \{\mathbf{s}_v\}_{v \in \mathcal{V}}) = \sum_{i=1}^m \Delta_{r_i} + \lambda_{ko} |\Upsilon|,
\end{equation}
where $\Upsilon = \{r\in \mathcal{R} \mid Asg_t(r) = \emptyset, \forall t\geq t_{req, r}\}$
is the set of unassigned requests,
and $\lambda_{ko} > 0$ is a penalty for not fulfilling a request.
The cost depends on the requests assigned to vehicles,
and the routes computed to complete them.

\subsection{Envy-Free Fairness}
\label{sec:envy-free-fairness}

The cost $J$ captures customer satisfaction (performance of the mobility-on-demand system).
However, it is equally important to consider fairness in allocating requests from the drivers' perspective.
We formalize the notion of utility for vehicles, 
and impose envy-free division~\cite{brams1996fair} of requests over finite time horizons.

Let $\Gamma_v \subseteq \mathcal{R}$ be set of requests completed by vehicle $v$.
The utility of vehicle $v$ with maximum capacity $Cap_v$ is
\begin{equation}
\label{eq:total-utility}
  U_v(\Gamma_v) = \sum_{t=0}^H (Cap_v - c_v(t)),
\end{equation}
and captures the utilization of $v$ over the time horizon $H$.

The request assignment over the time horizon $H$ is called \emph{envy-free} if
$U_v(\Gamma_v) \geq U_{v'}(\Gamma_{v'})$, for all $v, v' \in \mathcal{V}$.
Due to the sequential arrival of requests, we can not impose the
envy-free condition over the total utility in the time horizon $H$.
Instead, we investigate the slightly weaker condition that the vehicles'
utilities are envy-free when re-computing assignments at requests' arrival times.

\begin{problem}[Fair Request Assignment]
\label{pb:fair-assignment}
Given the set of vehicles $\mathcal{V}$ deployed in environment $(S, D, W)$,
and the set of requests $\mathcal{R} = \{r_1, \ldots, r_m\}$ arriving sequentially over time horizon $H$,
compute assignments $Asg_t$ at each time $t\in \{0, \ldots, H\}$ 
and routes $\mathbf{s}_v$ for all vehicles $v\in \mathcal{V}$
such that the vehicles' utilities satisfy the envy-free fairness conditions
and minimizes the cost $J$.
\end{problem}

\medskip
\noindent
{\bf Summary of the approach.}
When a new request arrives, or a vehicle becomes available,
We construct an assignment graph to match requests and vehicles.
The RTV graph has three layers (1) requests, (2) trips, and (3) vehicles.
Edges that connect vehicles to trips serving a subset of active requests
are computed via an automata-based routing procedure.
We construct product automata between the motion model (road network)
of a vehicle, and the DFAs corresponding to the requests.
The route is then computed via a shorted path method (e.g., Dijkstra algorithm)
applied on the product automaton graph and projection onto the motion model.
If maximum waiting and delay times constraints are met,
we add the edge to the assignment graph.
Lastly, we can formulate an ILP problem to minimize the sum of travel costs
such that the envy-free constraints hold for the allocated utilities of each vehicle.
The solution of the ILP provides the assignment scheme.

\section{Solution}
\label{sec:solution}

The mobility-on-demand ride-sharing problem can be translated to an ILP problem through the construction of a shareability graph and an assignment (RTV) graph~\cite{santi2014quantifying,alonso2017demand}.
Then we can apply graph search algorithms and provide efficient solutions.
This paper uses automata theory to construct the assignment graph,
and applies fair planning through envy-free constraints
and a proposed graph weight correction method.
In the following, we drop the time subscript $t$ to improve readability,
and whenever it is clear from context.

\begin{figure}
    \centering
    \includegraphics[width=0.9\linewidth]{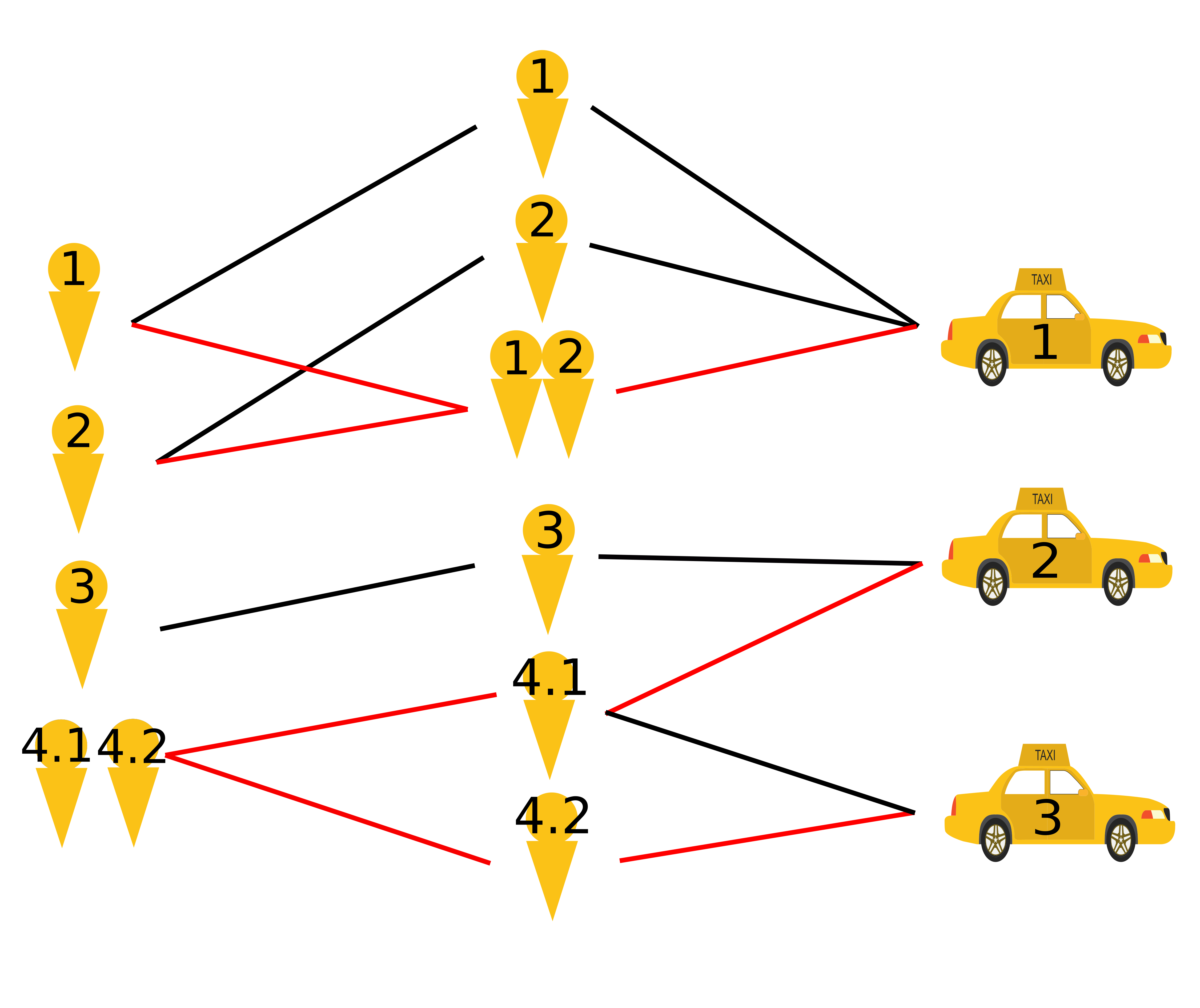}
    \caption{Example of RTV graph: The graph includes 4 requests and 3 vehicles. The fourth requests requires two vehicles. The red edges indicate a possible assignment scheme.}
    \label{fig:rtv}
\end{figure}

\begin{algorithm}[htb]
\caption{Fair Request Assignment Algorithm}
\label{alg:assignment}
\KwIn{$\mathcal{R}^a$ -- the active requests, $\mathcal{V}^a$ -- the available vehicles}
\KwOut{$Asg: \mathcal{V}^a \to \mathrm{Tr}$ -- vehicles to trips assignment}
\DontPrintSemicolon
\BlankLine
\small
\tcp*[l]{Construct RV Graph}
$\mathcal{G}^{RV} = (\mathcal{R}^a \cup \mathcal{V}^a, E^{RV} = \emptyset)$\;
\ForAll{$r, r' \in \mathcal{R}^a, r\neq r'$}{
  \lIf{$\mathrm{check\_share(r, r')}$}{
     $E^{RV} \gets E^{EV} \cup e(r, r')$
  }
}
\ForAll{$r \in \mathcal{R}^a$, $v \in \mathcal{V}^a$}{
  \lIf{$\mathrm{check\_trip(v, \{r\})}$}{
     $E^{RV} \gets E^{EV} \cup e(r, v)$
  }
}
\tcp*[l]{Construct RTV Graph}
$\mathrm{Tr}$ $\gets$ all the cliques of requests in $\mathcal{G}^{RV}$
that satisfy timing and capacity constraints\;
$\mathcal{G}^{RTV} = (\mathcal{R}^a \cup \mathrm{Tr} \cup \mathcal{V}^a, E^{RTV} = \emptyset)$\;
$E^{RTV} \gets \{(r, T) \mid r \in T\}$ \;
\ForAll{$T \in \mathrm{Tr}$, $v \in \mathcal{V}^a$}{
  \lIf{$\mathrm{check\_trip(v, T)}$}{
     $E^{RTV} \gets E^{EV} \cup e(T, v)$
  }
}

Asg = $\mathrm{solve\_ILP(\mathcal{G}^{RTV})}$\;
\Return Asg
\end{algorithm}

\subsection{Request-Trip-Vehicle (RTV) Graph}
\label{sec:rtv-graph}

The RTV graph batch assignment was introduced in~\cite{alonso2017demand}.
First, we construct the undirected Request-Vehicle (RV) graph
$\mathcal{G}^{RV} = (\mathcal{R}^a \cup \mathcal{V}^a, E^{RV})$ that captures requests that may be performed
by a vehicle in a single trip without violating the waiting time, and maximum delay constraints.
The RV graph's nodes are the requests and vehicles.
Edges $e(r, r')$ between two requests capture their shareability, i.e.,
can be served at the same time by a vehicle.
Edges $e(v, r)$ indicate whether vehicle $v$ can serve request $r$ under
the required timing constraints.

Next, we construct the undirected RTV graph $\mathcal{G}^{RTV}$ with nodes $\mathcal{R}^a \cup \mathrm{Tr} \cup \mathcal{V}$
and edges $E^{RTV}$, where $\mathrm{Tr}$ is the set of trips, see Fig.~\ref{fig:rtv}.
A trip $T_{v} \subseteq \mathcal{R}^a$ is a subset of requests serviced by a vehicle $v$.
Multiple vehicles $v_1, v_2, \ldots v_{n_r}$ may be needed to service a single request $r$,
in which case the $r$ is part of all their trips $T_{v_i}$, for all $i \in \{1, \ldots, n_r\}$.
Trips are formed from the RV graph by selecting its cliques~\cite{alonso2017demand} that satisfy
timing and capacity constraints for vehicles.
Thus, the RTV graph contains only potentially feasible trips of active requests for available vehicles.
Edges $e(r, T) \in E^{RTV}$ denote request $r$ is part of trip $T$,
while edges $e(T, v)$ denote that $v$ can serve requests in trip $T$, see Fig.~\ref{fig:rtv}.

Next, we define procedures to decide if edges $e(v, r)$ and $e(v, T)$
belong to the RV graph $\mathcal{G}^{RV}$ and the RTV graph $\mathcal{G}^{RTV}$, respectively,
for all $v \in \mathcal{V}^a$, $r \in \mathcal{R}$, and $T \in \mathrm{Tr}$.



\subsection{Automata-based Route Planning}

We construct product automata to obtain the RTV graph for scLTL requests and vehicles represented as a transition system (TS).
Formally, we have the TS $\TS_v=\left(S, s_{\text {init }}, D, W, \Pi, L\right)$
that captures vehicle $v$'s motion in the environment.
The set of propositions $\Pi$ includes the active requests' pick-up propositions $\pi_{pick, r}$.


\begin{definition}[Weighted product automaton at time $t$]
The weighted product automaton $\mathcal{P} = \mathcal{T} \otimes \mathcal{A}_{1} \otimes \ldots \otimes \mathcal{A}_{m}$ of vehicle $v$ at time $t$ is a tuple $\left(Q_{\mathcal{P}}, Q_{\text {init}, \mathcal{P}}, \delta_{\mathcal{P}}, F_{\mathcal{P}}, W_{\mathcal{P}}\right)$, where
\begin{itemize}
    \item $Q_{\mathcal{P}} = \{s, q_1, \cdots, q_m\}$;
    \item $Q_{init} = \{s_j, \pi_{pick,1}, \cdots, \pi_{pick,m} \}$, where $s_j$ is the current node of the $v_k$ in the map, i.e., $s_0 = s_{init}$; \\
    $q_{i, j}=
    \left\{\begin{array}{l}
        \delta_{i}\left(\pi_{pick, i}, L\left(s_{j}\right)\right)\text{ if } t_{pick, r_i} = t\\
        \delta_{i}\left(q_{i, j-1}, L\left(s_{j}\right)\right)\text{ if } t_{pick, r_i} < t\\
        q_{init, i} \text{ else, }
    \end{array}\right.$,\\
where $t_{pick,r_i}$ is the pick-up time for $r_i$;
\item $\delta_{\mathcal{P}} \subseteq Q_{\mathcal{P}} \times Q^{\prime}_{\mathcal{P}}$ is a transition function: \\
$\left(\left(s, q_{1}, \ldots, q_{m}\right),\left(s^{\prime}, q_{1}^{\prime}, \ldots, q_{m}^{\prime}\right)\right) \in \delta_{\mathcal{P}}$
if and only if $\left(s, s^{\prime}\right) \in R$\\ and $\left(q_{i}, L\left(s^{\prime}\right), q_{i}^{\prime}\right) \in \delta_{i}$;
\item $F_{\mathcal{P}}=\left\{\left(s, q_{1,k}, \ldots, q_{m,k}\right) \mid q_{i,k} \in F_{i}, \forall i \in \left\{1, \ldots, m\right\} \right\}$;
\item $ W_{\mathcal{P}}$: $\delta_{\mathcal{P}} \rightarrow \mathbb{R}_{+}$ is the weight function given by $W_{\mathcal{P}}(\left(\left(s, q_{1}, \ldots, q_{m}\right),\left(s^{\prime}, q_{1}^{\prime}, \ldots, q_{m}^{\prime}\right)\right)) = W(s, s^\prime)$.

\end{itemize}
\end{definition}

\subsubsection{Weighted product automaton for pairwise request-request in RV graph ($\mathrm{check\_share}$)}
This step checks if two requests $r$ and $r'$ can potentially be shared by the same vehicle, i.e., $\mathrm{check\_share}$ procedure used in Alg.~\ref{alg:assignment}.
Two requests can be combined pairwise if a virtual vehicle starting at one of their pick-up positions can complete both requests, i.e., satisfy the maximum delay and maximum wait time of both requests.
To achieve this, we construct a weighted automaton $\mathcal{P}_{RR} = \mathcal{T}_{virtual} \otimes \mathcal{A}_r \otimes \mathcal{A}_{r'}$ for $r$ and $r'$.
$\mathcal{T}_{virtual}$ is the transition system for the virtual vehicle with initial position $s_{init, virtual} \in \{q_{init,r}, q_{init,r'}\}$.
Then, we use a graph search method such as Dijkstra’s algorithm to check if an admissible path exists~\cite{sniedovich2006dijkstra}.
If it exists, edge $e(r, r')$ is added to the RV graph.
Moreover, these two requests are a potential candidate for a trip $T = \{r, r'\}$ denoted as edges $e(r, T)$ and $e(r', T)$ in the RTV graph.
For example, in Fig.~\ref{fig:small_road}, the trip $T_k = (r_1, r_2)$ with corresponding edges $e(r_1, T_k)$ and $e(r_2, T_k)$ is added to the RTV graph.

\subsubsection{Weighted product automaton for pairwise request-vehicle in RV graph ($\mathrm{check\_trip}$ with $|T| = 1$)}
The construction of the weighted product automaton for request-vehicle combination is similar to the product automaton for request-request.
For every available vehicle $v$ and request $r$, we construct a weighted product automaton $\mathcal{P}_{RV} = \mathcal{T}_v \otimes \mathcal{A}_r$.
The difference in this product automaton is that real-time vehicle information, i.e., the vehicle's position, is used.
Likewise, $v$ and $r$ are connected in the RV graph via edge $e(r, v)$ if an admissible path is found in the product automaton.
For example, edges $e(r_1, v_1)$ and $e(r_2, v_1)$ are in the RV graph for the case shown in Fig.~\ref{fig:small_road}.

\subsubsection{Weighted product automaton for RTV graph ($\mathrm{check\_trip}$ with $|T| > 1$)}
The connected requests and vehicles in the RV graph are feasible candidates for an assignment in the RTV graph with trips containing only one request, and the ride-sharing trips with more than one request can be built based on the RV graph. 
For requests $r$, $r'$ and vehicle $v_i$, if the pair $(r, r')$ is present in the RV graph, this means these two requests can share a vehicle, in the best-case scenario, when the vehicle is at their pick-up positions. And if $(r, v_i)$ and $(r', v_i)$ are also present in the RV graph, this means $v_i$ can serve $r$ or $r'$ under no sharing condition.
If both these conditions are satisfied, we can further validate the ride-sharing possibility of $v_i$ to serve both $r$ and $r'$ by constructing a weighted product automaton $\mathcal{P}_{RTV} = \mathcal{T}_{i} \otimes \mathcal{A}_r \otimes \mathcal{A}_{r'}$. 
And if an admissible path is found without violating the request constraints, $v_i$ and $T_j = \{ r$, $r'\}$ are grouped as a potential valid trip in the assignment and an edge $e(T_j, v_i)$ is created in the RTV graph to denote a potential assignment $(T_j, v_i)$.
For example $e(T_k, v_1), T_k = \{r_1, r_2\}$ would be created for Fig.~\ref{fig:small_road}. The RTV graph can be generated recursively for $Cap_v \geq 2$. In this paper, we consider the case where $Cap_v = 2$.

In the $\mathrm{check\_trip}$ step, for a vehicle $v_i$ and allocated trip $T_j = \{r_1, \cdots, r_n\}$,
the travel cost $\sigma_{v_i}(T_j)$ and travel utility $U_{v_i}(T_j)$ associated with each created edge are simultaneously generated as
\begin{equation}
    \begin{aligned}
    \sigma_{v_i}(T_j) &= \sum_{i=0}^n \Delta_{r_i},\\
    U_{v_i}(T_j) &= \sum_{t=0}^h (Cap_{v_i} - c_{v_i}(t)),
    \end{aligned}
\end{equation}%
where $n = |T_j|$ and $h$ is the trip serving duration.

\subsubsection{Weight Correction Based on History Utility}
One problem of non-fair assignment is that it doesn't consider the history utility, which may create a significant vacancy rate or disparity of total utility.
For example, during an off-peak hour, there might be a lesser number of requests than the number of vehicles available. 
Thus, some vehicles may not ever be allocated to any trips or only assigned with a low utility trip. Therefore, we make a cost correction based on a history utility to balance the accumulated utility over time. 

For a RTV graph, at any time step, the travel cost associated with edge $e(T_j, v_i)$ is adjusted in the following way
\begin{equation}
    \sigma_{v_i}^{new}(T_j) = \sigma_{v_i}^{old}(T_j)+ \alpha \cdot (U_{v_i} - U_{avg}),
\end{equation}
where $\alpha \in \mathbb{R}_{> 0}$ is a constant parameter and $U_{avg} = \sum_{i = 1}^p U_{v_i}/p,\, p = |\mathcal{V}|$ is the average history utility for all vehicles. 
After the weight correction, the traveling cost decreases for vehicles with low history utility and increases for vehicles with high history utility, thus favoring trips for vehicles with low history utility.

\subsection{ILP Formulation}
This section describes the $\mathrm{solve\_ILP}$ function in Algorithm~\ref{alg:assignment}.
We formulate the vehicle-sharing and fairness problem using
ILP, which needs to be updated when a new request arrives, or a vehicle becomes available.

A binary variable $\epsilon_{i,j} \in \{0, 1\}$ is introduced for each edge $e(T_j, v_i)$ in the RTV graph, $\epsilon_{i,j} = 1$ indicates that vehicle $v_i$ is assigned to trip $T_j$. In addition, a binary variable $\chi_{k} \in \{0, 1\}$ is introduced for each request $r_k$.
If $\chi_{k}$ takes the value one, it means that request $r_k$ is not served by any vehicles.

For multiple vehicles serving a single request, for compatibility with the bipartite graph representation, we divide the original request $\rho_i$ into $j$ sub-requests with each $\rho_{i, j} = 1$ and $\sum \rho_{i,j} = \rho_i$.
For example, $\phi_i$ is written as $\phi_{i,1} = \levent(\text{store 1})$ and $\phi_{i,2} = \levent(\text{store 2})$ with the same $t_{req}$.
Then, we constrain the assignment to contain either zero or all sub-requests. 

The objective of the ILP is to minimize the assignment cost.
The ILP formulation is defined as:
%
%
\begin{subequations}
\begin{align}
    \text{min } & \sum_{(i, j) \,:\, e(T_i,v_j)} \sigma_{v_i}(T_j)\, \epsilon_{i,j}  + \sum_{k=1}^m \lambda_{ko}\chi_{k},\label{a}\\
      \text{s.t.}\: & \sum_{i \, :\, e(R_k, T_i)} \sum_{j \, : \, e(T_i, v_j)} \epsilon_{i, j}+\chi_{k}=1, \quad \forall r_{k} \in \mathcal{R}, \label{b}\\
    &\sum_{i \, : \, e(T_i, v_j)} \epsilon_{i, j} \leq 1, \quad \forall v_{j} \in \mathcal{V},\label{c} \\
    & U_{v_i}(T_m) - \lambda \cdot U_{v_j}(T_n)  \geq M(\sigma_{v_i}(T_m) + \sigma_{v_j}(T_n) - 2) \label{d} \notag\\
\end{align}
\end{subequations}
The cost function defined in \eqref{a} minimizes the sum of travel cost plus a penalty $\lambda_{ko}$ for every unassigned request. \eqref{b} and \eqref{c} indicate each request is assigned to one vehicle at most, and each vehicle is assigned to one trip at most. 
The envy-free fairness constraint is captured by constraint in \eqref{d} using the \emph{big M} method, where $M$ is a constant value that is larger than the maximum value of all trip utilities.

Note that we do not need to compare the utilities of every pair of vehicles at a time step.
For example, $T_m$ may be infeasible for $v_j$ in ~\eqref{d}, thus lending the comparison trivial.
Moreover, some vehicles may not be available at $t$.

Note that in a strict envy-free allocation, the resulting matching may be undesirable in some scenarios.
For example, suppose in a road network containing two vehicles $v_1$ and $v_2$ and two requests $r_1$ and $r_2$,
both $v_1, v_2$ can serve $r_1$ and only $v_1$ can serve $r_2$.
Ideally, the optimal solution matching pair is $(v_1, r_2)$ and $(v_2, r_1)$.
However, if $r_1$ has larger utility than $r_2$ for vehicle $v_1$,
then the envy-free matching would only allocate $(v_1, r_2)$ leaving $r_1$ vacant and $v_2$ unoccupied.
Otherwise, if either vehicle serves $r_1$, then they would envy the other.

To tackle the problem, we adapt the relaxation idea \emph{envy-free up to one item}.
Two agents would not envy each other if one item is removed from the environment \cite{budish2011combinatorial}.
However, the assignment is indivisible and one vehicle can only be allocated to one assignment at a time, and as such removing one item is not feasible.
Thus, we introduce a variable $\lambda \in [0,1]$ to regulate the approximation of envy-free to~\eqref{d}.
When $\lambda = 0$ the envy-free constraints is disabled, and $\lambda = 1$ enforces strict envy-free.
In the previous example, if $U_{v_1}(r_2) \geq \lambda \cdot U_{v_2}(r_1)$ holds for
$\lambda \leq  \frac{U_{v_1}(r_2)}{U_{v_2}(r_1)}$,
the envy-free allocation is $(v_1, r_2)$ and $(v_2, r_1)$.

This approximately envy-free approach can still suffer from the previous issue in extreme scenarios, for example, when there is significant utility disparity between the two available requests, making the relaxation $\lambda$ be close to 0 to obtain the optimal overall utility. And because of a large number of requests and long working period, the total utility deviation among vehicles can be similar regardless of the envy-free enforced.


A greedy solution to maximize the serving rate while minimizing the travel cost is first computed as an initial guess for the ILP. Once solving the above ILP problem, the assignment scheme constructs the optimal path for vehicles as the accepted and shortest run in the corresponding product automaton projected onto the transition system.

\section{Simulation Results}
In this section, we present simulation results to demonstrate the performance in terms of scalability and fairness in a realistic road map.

\vspace{-2mm}

\subsection{Simulation Specifications}
We simulated the result in the mid-Manhattan map, which models the road intersections as nodes. The map contains 184 nodes, and the edges are weighted by real travel duration obtained from real taxi driving data. 
For details about the dataset, see~\cite{alonso2017demand}. 
Gurobi was used to solve the ILP \cite{optimizationinc}.
The simulation duration is set to 20 minutes with varying the number of vehicles and requests. The scLTL formulas were generated from the following scLTL pattern stochastically.  

\textbf{scLTL pattern:}
\vspace{-3mm}
\begin{equation*}
\begin{gathered}
    \begin{aligned}
    \tilde{\phi}_{1}\left(s_{pick}, s_{1}, s_{2}\right) &= \levent(s_{pick} \wedge \levent\left(s_{1} \wedge \levent\left(s_{2}\right)\right)),\\
    \tilde{\phi}_{2}\left(s_{pick}, s_{1}, s_{2}\right) &= \levent(s_{pick} \wedge\levent\left((s_{1} \vee s_{2}\right) \wedge s_{3})),\\
    \tilde{\phi}_{3}\left(s_{pick}, s_{1}, s_{2}, s_{3}\right) &= \levent(s_{pick} \wedge \levent\left(s_{1} \wedge  (s_{2} \vee s_{3}\right))),\\
    \tilde{\phi}_{4}\left(s_{pick}, s_{1}, s_{2}, \cdots, s_{n}\right) &=  \levent(s_{pick} \wedge\levent\left(s_{1} \wedge  (\neg s_{2} \wedge \cdots \neg s_{n}\right)),
    \end{aligned}
\end{gathered}
\end{equation*}
where $s_{i}$ are locations in the road map. The multi-vehicles serving requests are combinations of the above scLTL pattern echoing the sub-requests division technique. Furthermore, we also generate random pick-up positions and arrival times for each request in a uniform Poisson process. 
The maximum waiting time and delay time are set to $\Omega_{max}=2$ and $\Delta_{max} = 4$ minutes for every request, respectively.

The vehicle's transportation capacity is set to at most two requests at a time for ride-sharing, i.e., $Cap_v = 2$, and we generate the initial positions for the vehicles at time $t=0$ stochastically.
The envy-free variable $\lambda$ is set to 0.5.

\subsection{Simulation Results and Discussions}
We simulate the results by varying the vehicle-to-request ratio to demonstrate fairness and run time performance.

\begin{figure}[htb]
    \centering
    \includegraphics[width = 0.9\linewidth]{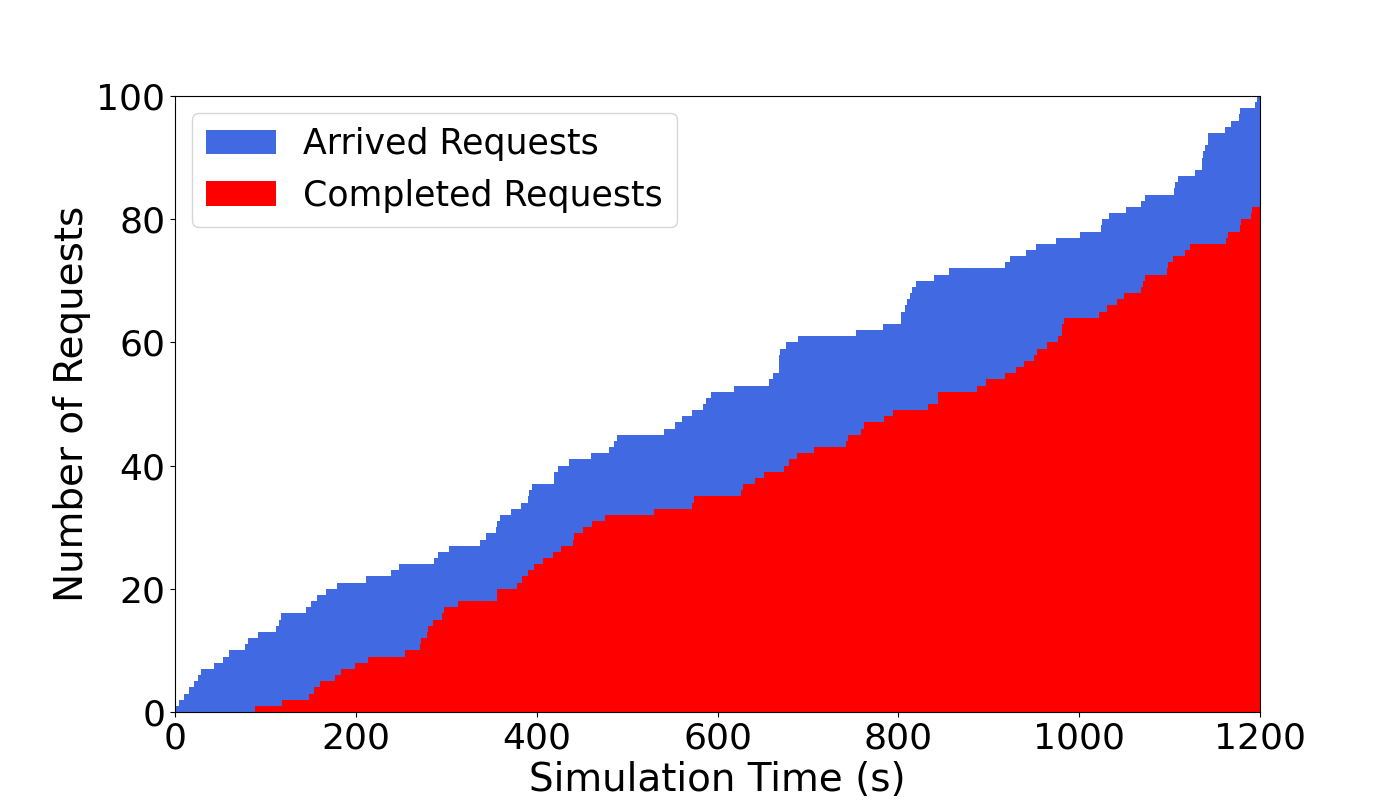}
    \caption{Number of arrived and completed requests over time}
    \label{fig:arrival}
\end{figure}

In the simulation shown in Fig.~\ref{fig:arrival} we consider 50 vehicles and 100 requests in the network. The figure shows the steady serving pace along with the gradually arriving requests.
 
\begin{figure}[htb]
    \centering
    \subfloat[\label{subfig:vacancy}]{%
       \includegraphics[width=0.9\linewidth]{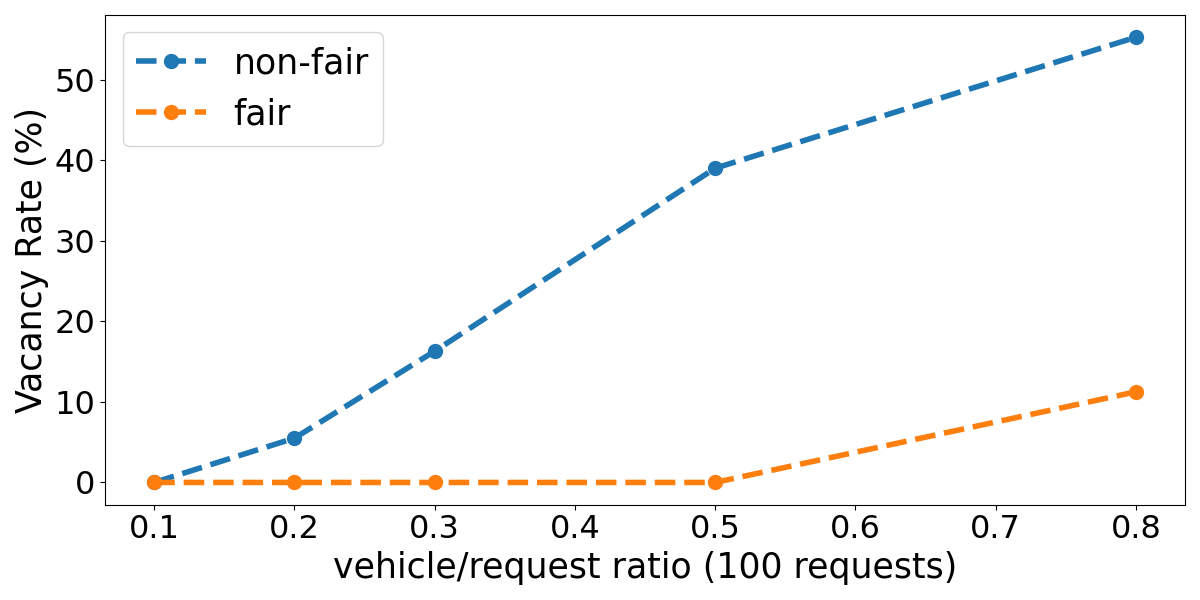}}
    \\
    \subfloat[\label{subfig:deviation}]{%
       \includegraphics[width=0.9\linewidth]{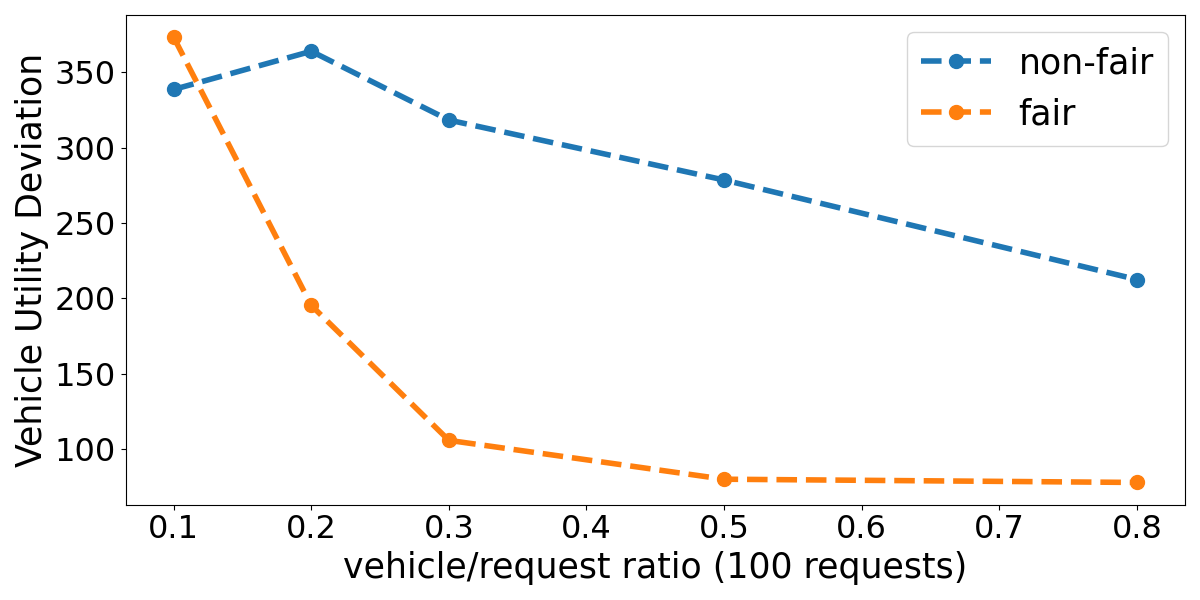}}
    \caption{Comparison between fair and non-fair planning. (a) Vehicle Vacancy Rate. (b) Vehicle Utility Deviation}
    \label{fig: compare}
\end{figure}

Fig.~\ref{fig: compare} shows the comparison between fair planning versus non-fair planning, i.e., plan without envy-free constraints and weight correction. 
Each data point in the figure is the average data of ten runs.
We study the vacancy rate and utility deviation as fair criteria.

The vacancy rate is defined as the percentage of unoccupied vehicles in the entire simulation time. Intuitively, the higher ratio of the vehicle to request, the higher the vacancy rate, as shown in Fig.~\ref{subfig:vacancy}.
Under fair planning conditions, the vacancy rate is significantly reduced.

In addition, Fig.~\ref{subfig:deviation} shows the history of utility deviation. The figure shows that fair planning also significantly decreases the utility deviation among vehicles.
The result of Fig.~\ref{fig: compare} is expected as the fair planning makes the ILP solution favor vehicles with low utility and, thus, reduces the vacancy rate as well.
In Fig.~\ref{subfig:deviation}, when there are small number of vehicles in the road, the utility deviation is similar.
This is also expected as almost every vehicle would receive an assignment once it becomes available, making fair planning similar to the non-fair baseline.

\begin{figure}[htb]
    \centering
    \subfloat[\label{subfig:requests}]{%
       \includegraphics[width=0.87\linewidth]{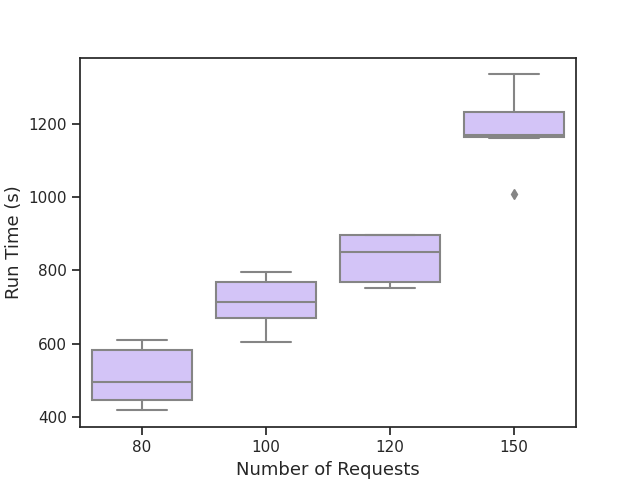}}
    \\
    \subfloat[\label{subfig:vehicle}]{%
       \includegraphics[width=0.87\linewidth]{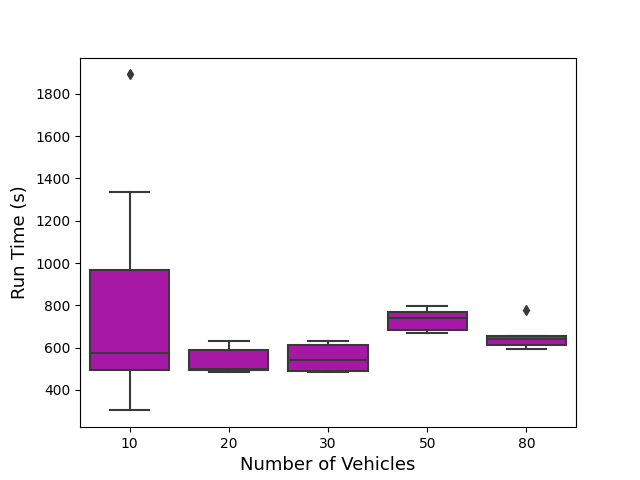}}
    \caption{Run Time Performance (a) for fixed number of 50 vehicles (b) for fixed number of 100 requests}
    \label{fig: Run Time}
\end{figure}

Fig.~\ref{fig: Run Time} shows the average computational run time performance for the simulation. We fix the number of vehicles and requests in Fig.~\ref{subfig:requests} and Fig.~\ref{subfig:vehicle} to demonstrate the scalability of our approach. 
The construction of the road map and RTV graph contribute most of the simulation time. The results show run time is similar for a fixed number of requests, and increases with the number of requests.

\section{Conclusions and Future Work}
A fair planning mobility-on-demand with temporal logic requests study is presented in this paper. The scLTL formulated requests allow passengers to define complex requests. 
We employ envy-free allocation and a utility-based weight correction to achieve a fair division of requests for vehicles.
We show that fair planning significantly decreases the vacancy rate and utility deviations between vehicles compared to a baseline that does not consider fairness constraints.
Moreover, we show that our method scales well with the number of vehicles and requests.



\bibliographystyle{ieeetr}
\bibliography{reference.bib}

\end{document}